\documentstyle[12pt]{article}

\input{psfig}

\def\cmm2{{\,\rm cm^{-2}}}
\def\cm2{{\,{\rm cm}^2}}
\def\cmm3{{\,{\rm cm}^{-3}}}
\def\gcmm3{{\,{\rm g\,cm^{-3}}}}

\def\mpl{{m_{\rm Pl}}}

\def\la{\mathrel{\mathpalette\fun <}}
\def\ga{\mathrel{\mathpalette\fun >}}
\def\fun#1#2{\lower3.6pt\vbox{\baselineskip0pt\lineskip.9pt
  \ialign{$\mathsurround=0pt#1\hfil##\hfil$\crcr#2\crcr\sim\crcr}}}

\begin{document}
\pagestyle{empty}
\begin{center}
\bigskip

\rightline{FERMILAB--Pub--97/010-A}
\rightline{CU-TP-831}
\rightline{IASSNS-HEP-97/47}
\rightline{hep-th/9705035}
\rightline{submitted to {\it Physical Review D}}

\vspace{1in}
{\Large \bf Pre-Big-Bang Inflation Requires \\
\bigskip Fine Tuning}
\bigskip

\vspace{.2in}
Michael S. Turner$^{1,2}$ and Erick J. Weinberg$^{2,3,4}$\\

\vspace{.2in}
{\it $^1$Departments of Astronomy \& Astrophysics and of Physics \\
Enrico Fermi Institute,
The University of Chicago, Chicago, IL~~60637-1433}\\

\vspace{0.1in}
{\it $^2$NASA/Fermilab Astrophysics Center\\
Fermi National Accelerator Laboratory, Batavia, IL~~60510-0500}\\

\vspace{0.1in}
{\it $^3$Department of Physics, Columbia University\\
New York, New York~~10027}

\vspace{0.1in}
{\it $^4$School of Natural Sciences, Institute for Advanced Study \\
Princeton, New Jersey~~08540}

\end{center}

\vspace{.3in}

\centerline{\bf Abstract}
\begin{quote}

The pre-big-bang cosmology inspired by superstring theories has been
suggested as an alternative to slow-roll inflation.  We analyze, in
both the Jordan and Einstein frames, the effect of spatial curvature
on this scenario and show that too much curvature --- of either sign
--- reduces the duration of the inflationary era to such an extent
that the flatness and horizon problems are not solved.  Hence, a
fine-tuning of initial conditions is required to obtain enough
inflation to solve the cosmological problems.

\end{quote}

\newpage
\pagestyle{plain}
\setcounter{page}{1}
\newpage

\section{Introduction}

The pre-big-bang cosmology inspired by superstring theories has been
suggested as a possible implementation of the inflationary-universe
scenario \cite{prebb}.  This cosmology is based on a spatially flat
solution in which the kinetic energy of a massless dilaton drives an
accelerated expansion toward a singularity at time $t_{\rm sing}$,
with the scale factor $a(t) \propto (t_{\rm sing}-t)^{-1/\sqrt{3}}$.
Before the singularity is reached, stringy and/or nonperturbative
effects bring an end to the inflationary phase and, by mechanisms that
are not yet completely understood, effectuate a transition to a
standard Friedmann-Robertson-Walker (FRW) epoch of decelerating
expansion with the dilaton fixed at its present value.

In previous investigations of this scenario considerable effort has
been focused on the details of the graceful exit from the inflationary
era \cite{exit}; it is still not at all clear that this can be done.
In this paper we will assume, for the sake of argument, that
mechanisms for accomplishing this actually exist and will concentrate
instead on the initial conditions.  These have received little
attention in previous discussions, in large part because the
flat-space solution on which these discussions have been based has an
inflationary epoch that extends infinitely far back in time.

The situation is quite different once one admits the possibility of
even a small amount of spatial curvature, as
generality
considerations certainly require.  Although pre-big-bang inflationary
solutions still arise, the inflationary epoch has a finite duration
which depends upon the initial curvature.  Furthermore, it does not even
appear that the pre-inflationary era can be extended arbitrarily far
back.  For a closed ($k=1$) universe this is prevented by the
existence of an initial singularity, while the open universe ($k=-1$)
solution, although remaining nonsingular, becomes increasing
implausible as $t \rightarrow -\infty$.  For both cases, as well as
for their limiting $k=0$ case, one is thus led to view the scenario as
beginning with the appearance (e.g., by a quantum fluctuation) of a
sufficiently large, smooth region at an initial time $t_0$, with the
subsequent evolution of this region being determined by the classical
field equations and the initial data at $t_0$.  We will find that, in
contrast \cite{nohair} to slow-roll inflation, the pre-big-bang
scenario is quite sensitive to these initial conditions.

In Sec.~2 we obtain the pre-big-bang solutions with arbitrary spatial
curvature, and discuss the requirements that must be satisfied in
order that there be enough inflation to solve the horizon and flatness
problems of the standard cosmology.  We then show how these
requirements can be phrased as constraints on initial conditions.  We
carry out this discussion in the Jordan frame, in which the Planck
mass is a time-dependent quantity depending on the dilaton field and
the fundamental string length $\ell_{\rm st}$ is fixed.  In Sec.~3 we
describe the somewhat different, but equivalent, picture that results
if one works in the Einstein frame, where the Planck mass is fixed and
$\ell_{\rm st}$ varies with time.  Section~4 contains some concluding
remarks.

\section{Pre-Big-bang Cosmology with Curvature}

\subsection{Flat-space solutions}

The evolution of the universe during the pre-big-bang phase of this
scenario is governed by the tree-level, low-energy effective action
\begin{equation}
    S_{\rm eff} = {1\over 2}\int d^4x \sqrt{-g}\, e^{-\sigma} \left[
    \ell_{\rm st}^{-2} (R + \partial_\mu \sigma \partial^\mu \sigma) 
    + R^2 + {\rm matter~terms} + O(e^\sigma) \right]\,.
\label{action}
\end{equation}
If we assume a Robertson-Walker metric with scale factor $a(t)$, the
Friedmann equation takes the form
\begin{equation}
    \left( H + {1\over 2} {\dot\Phi \over \Phi} \right)^2 =
     {1\over 12} \left({\dot \Phi \over \Phi} \right)^2 
    +  {8 \pi \over 3} {\rho \over \Phi} - {k \over a^2}\, ,
\label{friedmann}
\end{equation}
where a dot indicates differentiation with respect to
time, $H=\dot a/a$, and the Brans-Dicke field $\Phi$ and the associated
time-dependent Planck length $\ell_{\rm Pl}$ are related to the
dilaton field by
\begin{equation}
    \Phi =\ell_{\rm Pl}^{-2} = \ell_{\rm st}^{-2} e^{-\sigma}\,.
\end{equation}

   Equation~(\ref{friedmann}) must be supplemented by the equation of motion for
the dilaton field,
\begin{equation}
 {d \over dt}\left( \dot\Phi a^3 \right) 
    = 8\pi \left({\rho-3p\over a^3}\right) \, ,
\label{dilatoneq}
\end{equation}
as well as the equations governing the fields responsible for the
energy density $\rho$.  If the latter is entirely due to radiation, as
we assume henceforth, $\rho \propto 1/a^4$ and the right hand side
of Eq.~(\ref{dilatoneq}) vanishes, implying that
\begin{equation}
       B = -\dot\Phi a^3  
\label{Bdef}
\end{equation}
is a constant.  In order to obtain a solution in which the string 
coupling $e^\sigma$
evolves from weak to strong, we require that $B$ be positive.
Using these results, we may rewrite the Friedmann equation as 
\begin{equation}
    \left( \dot a - {B \over 2a^2\Phi} \right)^2 =
     {B^2\over 12(a^2 \Phi)^2 } 
    +  {bB \over \sqrt{3}\,a^2\Phi} - k \,,
\label{friedtwo}
\end{equation}
where
\begin{equation}
      b \equiv {8\pi \sqrt{3} \over 3 B} \rho_{\rm rad} a^4
\end{equation}
is a constant.

    The solution of these equations is particularly simple when
$k=b=0$.  With $\dot\Phi >0$ there are two solutions, with
$a(t) \sim (\pm t - {\rm const})^{\pm 1/\sqrt{3}}$.  The lower signs
give the inflationary solution underlying the pre-big-bang scenario;
this solution may be written as
\begin{eqnarray}
    a(t) &=& A  (t_{\rm sing}-t)^{-1/\sqrt{3}} \,,
                  \nonumber\\
    \Phi(t) &= &  {BA^{-3} \over 1+\sqrt{3}}
         (t_{\rm sing}-t )^{1+\sqrt{3}}      
\label{flatcase}
\end{eqnarray}
with $t_{\rm sing}$ and $A$ being arbitrary constants.  If
$k=0$, the freedom to make an overall time-independent rescaling
of $a$ means that $B$ has no invariant meaning.  For the cases with
nonzero spatial curvature we will eliminate this freedom by adopting
the convention that $|k|=1$, so that $a$ is the curvature radius of space.

    As we will see, this flat-space solution is also a good
approximation to the final, inflationary stages of the $k=\pm 1$
solutions.  We may view the inflationary era of these solutions as
beginning at a time $t_i$, when they are well approximated by
Eq.~(\ref{flatcase}), and ending at a time $t_f < t_{\rm sing}$ when
the solution ceases to be reliable, either because the the coupling
$e^\sigma$ has become large enough that higher-order loop corrections
to the low-energy effective action can no longer be neglected or
because the space-time curvature has become so great that stringy
and/or quantum gravity effects are significant.  The amount of
inflation during the interval between these can be measured by the
factor $Z$ by which the comoving Hubble length $(Ha)^{-1}$ decreases.
Using Eq.~(\ref{flatcase}), we find that
\begin{equation}
       Z =  { H(t_f) a(t_f) \over H(t_i)a(t_i) } 
         = \left({ \Phi(t_i) \over \Phi(t_f)}\right)^{1/\sqrt{3}}
         = \left({t_{\rm sing}-t_i \over t_{\rm sing}-t_f }\right)
              ^{(1+\sqrt{3})/\sqrt{3}} \,.
\label{firstZeq}
\end{equation}
If the presently observed universe is contained within a region that
had a size $H^{-1}(t_i)$ at time $t_i$, then solution of the horizon
problem requires that $Z > e^{60}$ (see e.g., Ref.~\cite{kt}).
Hence,  the effective Planck mass must change by
a very large amount, $\sqrt{\Phi (t_i)/\Phi (t_f)} \ga e^{104}$ and,  
because the growth of the scale factor is a power-law and
not exponential, the inflationary period must be of
long duration, $(t_{\rm sing}-t_i)/(t_{\rm sing} -t_f) \ga
e^{38}$.

Since inflation ends if the coupling becomes strong, $\Phi(t_f)
\ga \ell_{\rm st}^{-2}$.  Similarly, the fact that the classical
equations are reliable only if the curvature is less than the string
scale implies that $H^{-1}(t_f) \sim (t_{\rm sing} -t_f) \ga \ell_{\rm
st}$.  Taken together, these inequalities imply the bound
\begin{equation}
      Z  \la {\rm Min} \, \left\{
        \left( \Phi(t_i)  \ell_{\rm st}^2 \right)^{1/\sqrt{3}} \, , 
      \left({t_{\rm sing}-t_i \over \ell_{\rm st} }\right)
              ^{(1+\sqrt{3})/\sqrt{3}}   \right\} \,.
\label{firstZbound}
\end{equation}

\subsection{Solutions with spatial curvature}

    Let us now turn to the solutions with nonzero spatial
curvature.\footnote{For another treatment of these equations, see
Ref.~\cite{lf}; some related solutions involving dilaton fields in
models with spatial curvature  are discussed in
Ref.~\cite{curvedsolutions}.} 
We begin by solving Eq.~(\ref{friedtwo}) for $\dot a$ and then
substituting that result in Eq.~(\ref{Bdef}).  Introducing the
variable
\begin{equation}
    \psi = {\sqrt{12} \over B} a^2 \Phi \, ,
\label{psidef}
\end{equation}
which is proportional to the square of the Einstein-frame scale factor,
we can write the resulting equations as 
\begin{eqnarray}
    \dot a &=& {1\over \psi} \left[ \sqrt{3} 
    \pm \sqrt{1+ 2b\psi -k\psi^2 } \right] \,,
\label{adoteq} \\
   \dot \psi &=&  \pm {2\over a} \sqrt{1 +2b\psi -k \psi^2} \,.
\label{psidoteq}
\end{eqnarray}
The choice of signs in these equations is determined by the
requirement that $\psi \sim a^2\Phi$ eventually tend toward zero, so
that the curvature term will become negligible and the solution can
approach the inflationary $k=0$ solution (\ref{flatcase}).  For $k=-1$
the sign of $\dot\psi$ can never change, and so the lower signs in
Eqs.~(\ref{adoteq}) and (\ref{psidoteq}) must be chosen throughout.
For $k=1$, where $\dot\psi$ changes sign, the upper
signs apply at early times and the lower ones at late times.

    Rather than solve these equations directly, we introduce a
parameter $\eta$ that satisfies
\begin{equation}
    \dot \eta = {1\over a} \,.
\label{etadot}
\end{equation}
Examining Eq.~(\ref{psidoteq}), we see that $d\psi/d\eta = \dot\psi/
 \dot\eta$ depends only on $\psi$.  Hence, $\psi$ can be obtained as a
function of $\eta$ by straightforward integration.  After this result
is substituted into Eq.~(\ref{adoteq}), a second integration yields
$a(\eta)$.  The solutions thus obtained are
\begin{eqnarray}
    a (\eta ) &=& 3^{-1/4} \sqrt{B} \, Q^{-1}
          [ C(\eta) - bS(\eta)]^{(1+\sqrt{3})/2}  
          [-S(\eta)]^{(1-\sqrt{3})/2} \,,   \nonumber \\
    \Phi (\eta ) &=&  Q^2
        \left[ {-S(\eta) \over C(\eta) - bS(\eta)} \right]^{\sqrt{3}}\,,
\label{fullsoln}
\end{eqnarray}
with $Q$ an arbitrary constant.  For $k=-1$, $C(\eta)$ and $S(\eta)$
denote $\cosh\eta$ and $\sinh\eta$, respectively, while for $k=1$ they
denote $\cos\eta$ and $\sin\eta$. 

    For small negative values of $\eta$, both the $k=1$ and the $k=-1$
solutions approach the $k=b=0$ solution (\ref{flatcase}), with
$\eta=0$ corresponding to $t_{\rm sing}$.  At earlier times, however,
the behavior of these solutions is quite different.  For the moment,
we concentrate on the case $b \la O(1)$, for which radiation is never
dominant.

\begin{figure}[t]
\centerline{\psfig{figure=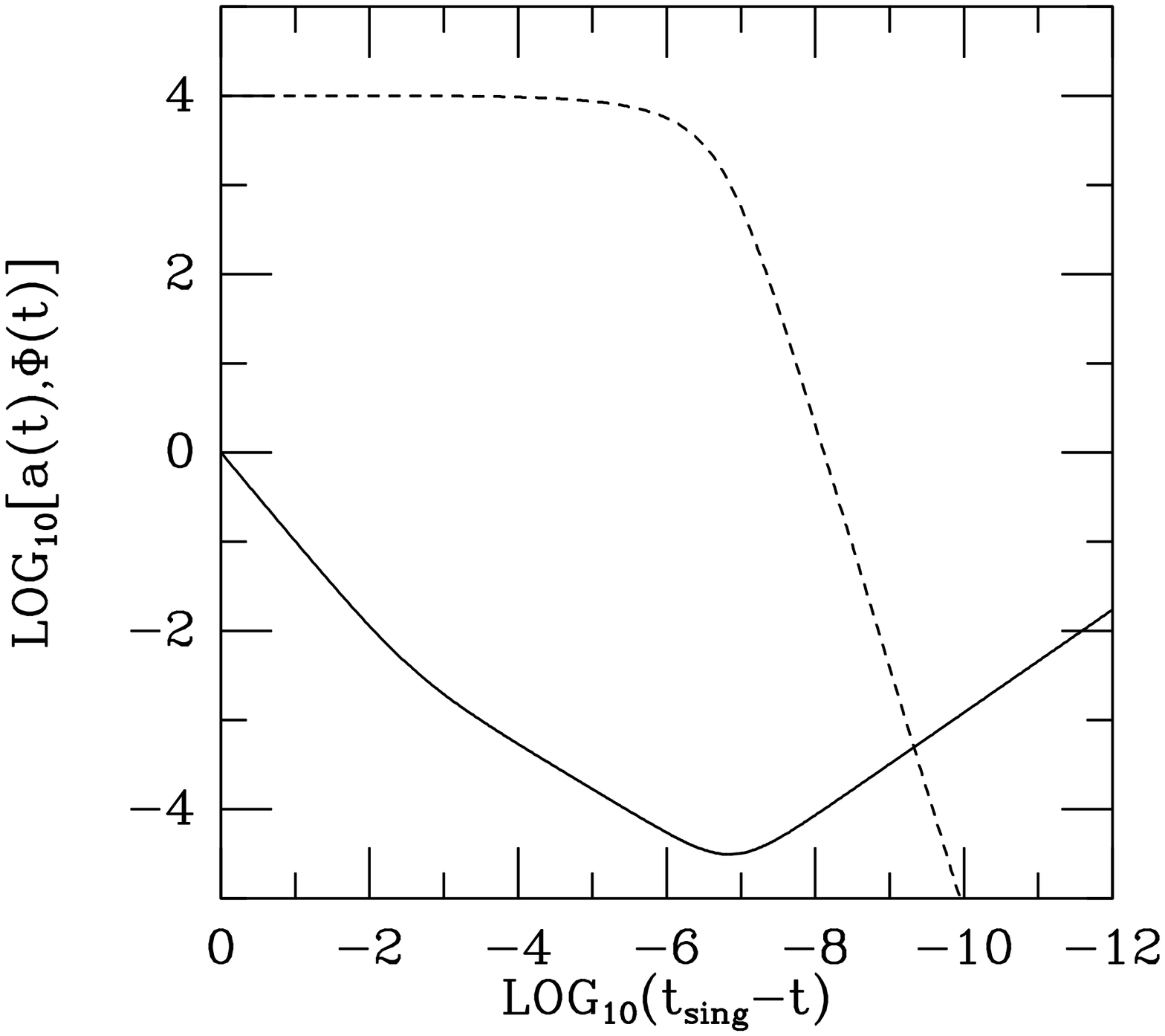,width=4.5in}}
\caption{Evolution of the cosmic scale factor $a(t)$
(solid curve) and the Brans-Dicke field $\Phi (t)$ (broken
curve) for $k=-1$ and $b\gg 1$; the scales for $a$ and
$\Phi$ are arbitrary.  Curvature-dominated and
radiation-dominated phases precede the inflationary phase;
the Brans-Dicke field remains constant during the
pre-inflationary phases.  For the case of $b\ll 1$ there is no
intermediate radiation-dominated phase and hence no change of slope in
the contracting portion of the $a(t)$ curve.}
\label{fig:kminus1}
\end{figure}

\begin{figure}[t]
\centerline{\psfig{figure=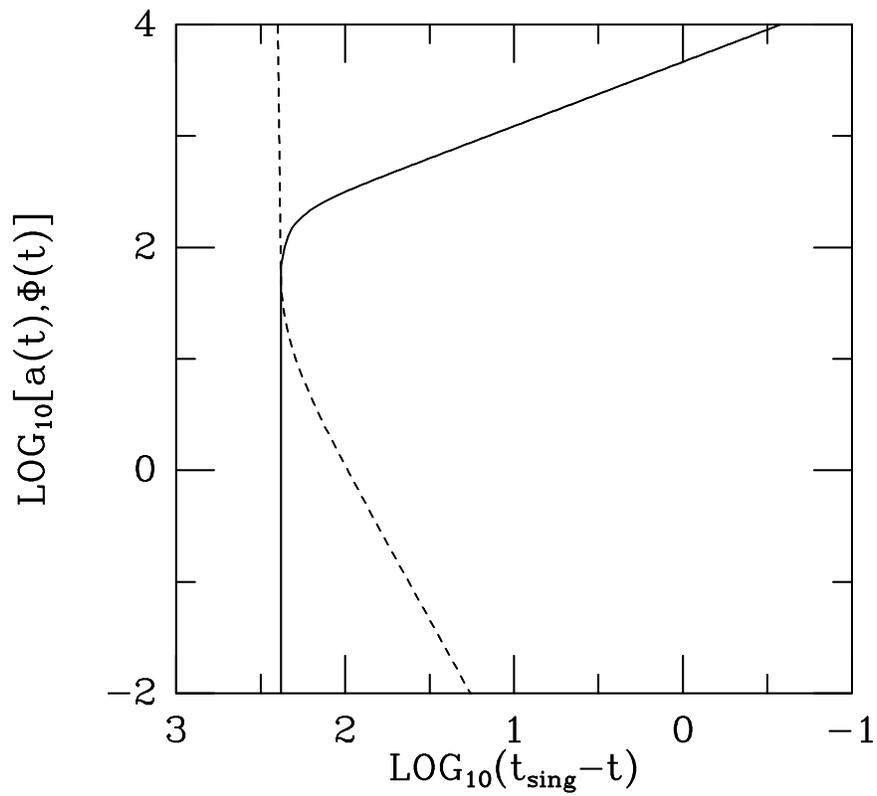,width=4.5in}}
\caption{Same as Fig.~\protect\ref{fig:kminus1} for $k=1$ and
$b=0$.  An initial singularity precedes
the inflationary phase (both the scale
factor and the Brans-Dicke field are singular).}
\label{fig:kplus1b0}
\end{figure}

    The $k=-1$ solution (see Fig.~\ref{fig:kminus1})
remains nonsingular for all
$\eta <0$.  At very early times (large negative values of $\eta$), $a$
decreases linearly with time, while the dilaton field remains
approximately constant at a value $\Phi(-\infty) \sim Q^2$.  The scale
factor reaches a minimum $a_{\rm min}$ when $-\eta$ is of order unity,
and then begins to grow as the universe goes over into the
dilaton-dominated inflationary epoch.  By contrast, the $k=1$ solution
(Fig.~\ref{fig:kplus1b0}) has an initial singularity,
with vanishing $a$ and diverging
$\Phi$, a finite time $t_{\rm total}$ before the final
singularity. 

   In either case, the solution begins to approximate the inflationary
flat space solution when $-\eta$ is of order unity, implying 
that $\Phi(t_i) \equiv [\ell_{\rm Pl}(t_i)]^{-2} \sim Q^2$ and hence
that  
\begin{equation}
    a(t_i ) \sim a_{\rm min} \sim \sqrt{B} \, \ell_{\rm Pl}(t_i)\,,
\label{aieq}
\end{equation}    
and [from integration of  Eq.(\ref{etadot})]
\begin{equation}
    t_{\rm sing} -t_i \sim {1\over 2} t_{\rm total}
         \sim \sqrt{B} \, \ell_{\rm Pl}(t_i)\,.
\end{equation}    
Substituting these results into Eq.~(\ref{firstZbound}) and using
Eq.~(\ref{aieq}), we find that
\begin{equation}
     Z  \la {\rm Min} \, \left\{
        \left( \Phi(t_i)  \ell_{\rm st}^2 \right)^{1/\sqrt{3}} \, , 
      \left({B \over \Phi (t_i) \ell^2_{\rm st} }\right)
              ^{(1+\sqrt{3})/(2\sqrt{3})}  \right\}   < B^{1/3}  
            \sim   \left({ a(t_i) \over
        \ell_{\rm Pl}(t_i)} \right)^{2/3} \, .
\label{secondZbound}
\end{equation}

\begin{figure}[t]
\centerline{\psfig{figure=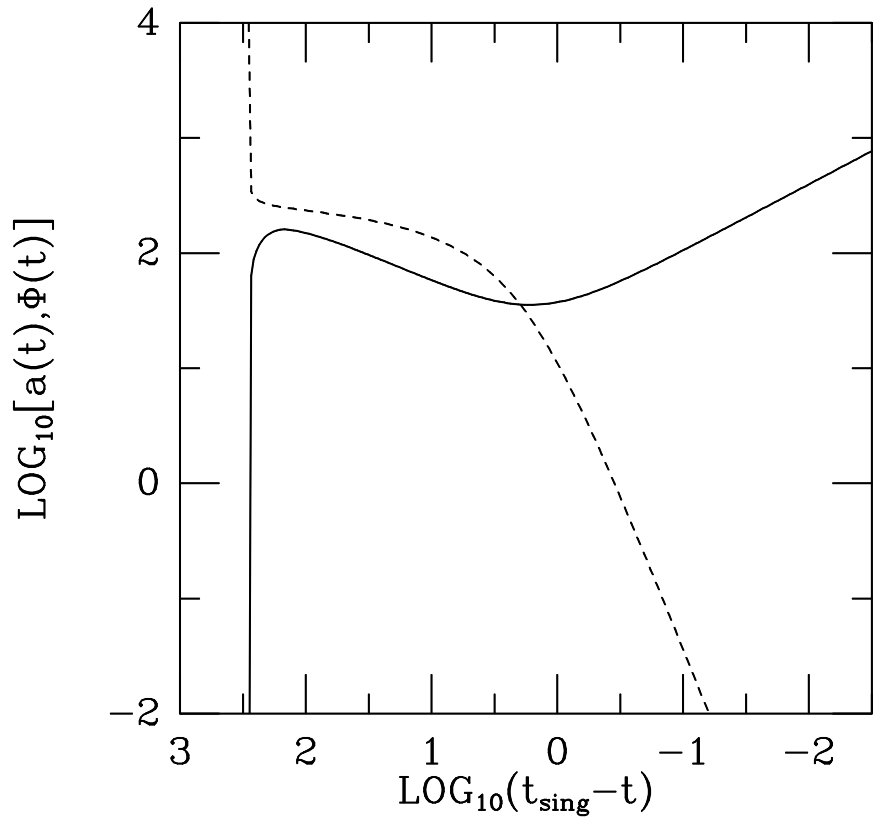,width=4.5in}}
\caption{Same as Fig.~\protect\ref{fig:kplus1b0} for $b\gg 1$.
In this case, the initial singularity is followed by
a radiation-dominated phase during which the scale
factor decreases and the Brans-Dicke field is
approximately constant.}
\label{fig:kplus1bgt0}
\end{figure}

    For $b \gg 1$, the $k=-1$ solution (Fig.~\ref{fig:kminus1})
    is qualitatively
rather similar, with $\Phi$ again remaining essentially constant until
the onset of inflation, but with the scale factor decreasing
as $(C-t)^{1/2}$.  On the other hand, the $k=1$ solution
(Fig.~\ref{fig:kplus1bgt0}) is quite different,
with a period of radiation-dominated
expansion and contraction (at a roughly constant value of $\Phi$)
preceding the final inflationary expansion.  In either case,
dilaton-dominated inflation begins when $-\eta \sim 1/b$, with
$\Phi(t_i) \sim Q^2/ b^{\sqrt{3}}$.  One finds that
\begin{eqnarray}
    a(t_i ) &\sim & b^{-1/2} \sqrt{B} \, \ell_{\rm Pl}(t_i) \,,
               \nonumber \\
    t_{\rm sing} -t_i &\sim&  b^{-3/2} \sqrt{B}  \, \ell_{\rm Pl}(t_i) \,,
\end{eqnarray}
and (for $k=1$) 
\begin{equation}
    t_{\rm total} \sim  a_{\rm max} \sim
         b^{1/2}  \sqrt{B} \, \ell_{\rm Pl}(t_i) \,,
\end{equation}
while Eq.~(\ref{secondZbound}) is replaced by  
\begin{equation}
     Z  \la  B^{1/3} b^{-1}    \sim  \left({ a(t_i) \over
       b\, \ell_{\rm Pl}(t_i)} \right)^{2/3}  \sim
   \left({t_{\rm sing} -t_i \over  \ell_{\rm Pl}(t_i)} \right)^{2/3} \,.
\label{binflationbound}
\end{equation}

\subsection{Constraints on initial conditions}

    The $k=0$ solution, Eq.~(\ref{flatcase}), can be extended
indefinitely far back into the past without encountering either a
singularity or a point where the underlying physical assumptions
clearly break down (in contrast, say, to the radiation-dominated FRW
solutions, which clearly cannot be trusted once the temperature
approaches the Planck scale).  Because of this, one can to a certain
degree sidestep the question of initial conditions.

    This is no longer the case once there
is spatial curvature.  For $k=1$, where there is an
initial singularity, this is obvious.  For $k=-1$ there is neither an
initial singularity nor a fixed time at which the solution must break
down.  However, as $t \rightarrow -\infty$ the physical size of the
region corresponding to the presently observed universe diverges.
Unless we want to have an infinitely large homogeneous region in the
far past, we must assume the appearance (e.g., by some stringy or
quantum gravitational mechanism) at some initial time $t_0$ of a
smooth region that is sufficiently homogeneous and isotropic to be
described by a Robertson-Walker metric.  The subsequent development of
this region will be determined by the initial values\footnote{There
is also a discrete choice for $\dot a_0$
corresponding to the sign ambiguity in Eqs.~(\ref{adoteq}) and
(\ref{psidoteq}); we will assume that the value corresponding to the
inflationary solution is chosen.} $a_0$, $\Phi_0$, and $\dot\Phi_0$.
The bounds obtained in the previous subsections place constraints on
these initial conditions.

    We consider separately the cases where inflation does not begin
until some time after $t_0$ and that where it begins immediately at
$t_0$; these correspond to $a_0 |\dot\Phi_0|/\Phi_0$ being less
than or greater than ${\rm Max}(1,b)$, respectively.  In the former
case, the relation between $Z$ and $B$ in Eqs.~(\ref{secondZbound})
and (\ref{binflationbound}) implies that
\begin{equation}
    Z\la {(-a_0^3 \dot\Phi_0)^{1/3}\over {\rm Max}(1,b)} \,.
\label{a0bound}
\end{equation}

   If, instead, inflation begins immediately at $t_0$,
Eq.~(\ref{firstZeq}) for $Z$ must be corrected to take into account
the fact that the portion of the classical solution in the interval
between $t_i$ and $t_0$ does not correspond to actually realized
inflation; the actual amount of inflation (i.e., from $t_0$ until
$t_f$) is reduced by a factor of
\begin{equation}
    {a^2_0 \Phi_0 \over a^2(t_i) \Phi(t_i) } 
    \sim { {\rm Max}(1,b) \over |{\dot\Phi_0}|/a_0\Phi_0 } \,.
\end{equation}
The bound corresponding to Eq.~(\ref{a0bound}) for this
case is
\begin{equation}
  Z\la \left( {\Phi_0^3 \over \dot\Phi^2_0}\right)^{1/3} \,.
\label{a0bound2}
\end{equation}
(The fact that $a_0$ does not explicitly enter the bounds in this case was
to be expected, since in the inflationary epoch the curved space
solutions approximate the flat space solution, for which there is an
overall scale ambiguity in the definition of $a$.)

   Furthermore by combining the relation between $Z$ and
$\Phi(t_i)$ in Eq.~(\ref{firstZbound}) with the fact that $\Phi$ is
monotonically decreasing, one obtains the bound 
\begin{equation}
    Z \la  (\Phi_0 \ell_{\rm st}^2)^{1/\sqrt{3}} \, ;
\end{equation}
in fact, this bound can be strengthened somewhat for certain choices
of parameters with $k=1$.

These bounds on the amount of inflation, Eqs.~(\ref{a0bound},
\ref{a0bound2}), make clear the sensitivity of pre-big-bang inflation
to initial conditions.  In the case where inflation does not begin
until sometime after $t=t_0$, either an increase in the initial
curvature, i.e., decreasing $a_0$, or an increase in the amount of
radiation, i.e., larger $b$, with all other quantities held fixed, can
reduce $Z$ to the point where it is insufficient to solve the
horizon and flatness problems.  In the other case, when inflation begins
immediately at $t=t_0$, an increase in ${\dot\Phi}_0$ or a decrease in
$\Phi_0$ can defeat inflation.

Recently, Veneziano has studied another, not unrelated, aspect -- the
effect of initial inhomogeneity and anisotropy on pre-big-bang
inflation \cite{veryrecent}.  Making the assumption of small initial
curvature, he showed that small amounts of inhomogeneity and
anisotropy do not prevent the ultimate transition to the pre-big-bang
inflationary phase and concluded that pre-big-bang inflation is
robust.  The first statement is consistent with our results -- we find
that radiation and spatial curvature only {\it postpone} the
inflationary phase -- and extends these to small levels of anisotropy
and inhomogeneity.  However, our interpretation is less rosy than
Veneziano's.  Since the end of pre-big-bang inflation is fixed by
other considerations, postponing the onset of the dilaton-dominated
phase can severely limit the beneficial effects of pre-big-bang
inflation, and we speculate that anisotropy and inhomogeneity may also be
able to defeat pre-big-bang inflation by postponing the onset of the
dilaton-dominated phase.

\section{The View from the Einstein Frame}

The theory defined by action of Eq. (\ref{action})
can be recast in a number of different, but equivalent,
forms by conformal rescalings of the metric.  In particular, the
conformal transformation ${\tilde g}_{\mu\nu}
= (\Phi/ \mpl^2) g_{\mu\nu}$ [and
hence $\tilde a/a = d{\tilde t}/dt= \Phi^{1/2}/\mpl $] takes the the
Jordan frame description that we have used thus far into the Einstein
frame description in which the gravitational part of the action takes
the standard Einstein-Hilbert form.  

The behavior in this frame is qualitatively quite different than that
in the Jordan frame.  For $k=-1$, $\tilde a$ (which, as noted earlier,
is proportional to $\sqrt{\psi}$) decreases monotonically, while for
$k=1$, $\tilde a$ vanishes at both the initial and final
singularities, with only a single maximum in between, no matter what
the value of $b$.  The inflationary epoch itself, instead of being a
period of accelerated expansion, is one of ever more rapid
contraction, with ${\tilde a}(t) \propto (\tilde t_{\rm sing} -\tilde
t)^{1/3}$, so that $\dot{\tilde a}$ and $\ddot{\tilde a}$ are both
negative.  Of course, the answers to physical questions cannot be
changed by a field redefinition, so if the horizon problem is solved
in one frame it must be solved in the other \cite{gv}.  Indeed, the
ratio in Eq.~(\ref{firstZeq}) that we used to characterize the amount of
inflation is the same in either frame:
\begin{equation}
        { H(t_f) a(t_f) \over H(t_i)a(t_i) } 
         = \left({t_{\rm sing}-t_i \over t_{\rm sing}-t_f }\right)
              ^{(1+\sqrt{3})/\sqrt{3}}        
         = \left({\tilde t_{\rm sing}-\tilde t_i \over 
          \tilde t_{\rm sing}-\tilde t_f }\right)^{2/3}
        = { \tilde  H(t_f)  \tilde a(t_f) 
                \over  \tilde H(t_i) \tilde a(t_i) } \,.
\end{equation}
Here, we have used the fact that during inflation ${\tilde H}{\tilde a}
\propto 1/{\tilde a}^2$.

In the Einstein frame, the sensitivity of the amount of inflation to
the amount of curvature (or radiation) can be understood as follows.
The terms in the Friedmann equation corresponding to curvature,
radiation energy density, and Brans-Dicke-field energy density vary as
$1/{\tilde a}^2$, $1/{\tilde a}^4$, and $1/{\tilde a}^6$,
respectively.  Inflation only begins when $\tilde a$ has become small
enough that the last of these terms is dominant.  Hence, by increasing
the curvature or the amount of radiation,
the duration of inflation is made shorter.

Although the two frames are mathematically equivalent, they do suggest
different levels of ``naturalness.''  The initial-condition
constraints that we obtained in the previous section arise (in
different forms) in both frames.  However, in the Einstein frame the
picture of inflation is far less compelling: A big, smooth region
emerges at the end of inflation because an even bigger smooth region
was present at the beginning of inflation.  These considerations can
be rephrased in terms of fundamental length scales.  If one views the
dilaton theory as being an effective theory based on an underlying
string theory with a fundamental length scale $\ell_{\rm st}$, then
the Jordan-frame picture, with inflation taking a small smooth region
into a large smooth one, is perhaps more natural.  However, nothing
that we have done has relied on any underlying string physics; a
similar scenario (up to the implementation of the graceful exit) could
be obtained from any generalized Brans-Dicke theory.\footnote{Some of
the difficulties associated with achieving sufficient inflation in
such theories have been discussed in \cite{janna}.}  In
this latter case, there is no fundamental length.  The only natural
way to characterize distances as ``large'' or ``small'' is relative to
the time-dependent Planck length.  But this leads immediately to a
difficulty.  The bounds in Eqs.~(\ref{secondZbound}) and
(\ref{binflationbound}) 
tell us that significant inflation during the
dilaton-dominated era is possible only in a universe whose
characteristic cosmological scales --- $a(t_i)$, $(t_{\rm sing}
-t_i)$, $a_{\rm min}$ [for $k=-1$], and $t_{\rm total}$ [for $k=1$]
--- are all enormous when measured in units of its characteristic
Planck length $\ell_{\rm Pl}(t_i)$.  The flatness problem is just such
a mismatch in scales.  Thus, the price of curing one set of
naturalness problems is the re-introduction of an earlier naturalness problem.

\section{Concluding Remarks}

One measure of the naturalness of a cosmological scenario is its
sensitivity to initial conditions.  Indeed, the primary motivation for
the inflationary paradigm was to solve the naturalness problems that
arose because the standard cosmology appeared to require a
finely tuned initial state.  A characteristic of previous
implementations of inflation is that they are
rather insensitive to the initial curvature \cite{nohair,st}.  For
example, in slow-roll inflation
\begin{equation}
\ln Z = {8\pi \over \mpl^2}\,\int_{\phi_i}^{\phi_{\rm end}} 
    {V(\phi )d\phi \over V^\prime (\phi )},
\end{equation}
is determined by the shape of the inflationary potential, the initial
value $\phi_i$ of the inflaton field, and the value $\phi_{\rm end}$
of the inflaton at which the slow-roll approximation breaks down; the
only constraint on the initial curvature is that it not have such a
large positive value that recollapse would occur before inflation
could commence.  Viewed in the context of inhomogeneous cosmological
models, this means that all sufficiently-large regions of space with
either negative or not-too-large positive curvature will inflate
\cite{ssj,star}.

In pre-big-bang inflation the end of the inflationary era is fixed,
while its beginning is delayed by curvature. Too much curvature
--- of either sign --- shortens the duration of the
inflationary era to the point that the flatness and horizon problems
are not solved.  Thus, pre-big-bang inflation requires fine-tuning of
initial conditions to solve these cosmological problems.  This
makes it less robust, and therefore less attractive as an
implementation of the inflationary paradigm.

\paragraph{Acknowledgments.}  We thank Janna Levin and other
participants of the 1995 Aspen Center for Physics Workshop
on Inflation for valuable
discussions.  This work was supported by the DoE (at Chicago, Columbia,
and Fermilab) and by NASA through grant NAG 5-2788 at Fermilab.

\newpage

\end{document}